# Dreams, Endocannabinoids and Itinerant Dynamics in Neural Networks: elaborating Crick-Mitchison's Unlearning Hypothesis

*Osame Kinouchi*
*Departamento de Física e Matemática*
*Faculdade de Filosofia, Ciências e Letras de Ribeirão Preto*
*Universidade de São Paulo*
*Avenida dos Bandeirantes 3900,*
*14040-901, Ribeirão Preto, SP, Brazil.*
*E-mail: osame@ffclrp.usp.br*

*Renato Rodrigues Kinouchi*
*Centro de Ciências Naturais e Humanas*
*Universidade Federal do ABC*
*Rua Santa Adélia, 166, Bangu*
*09210-170,  Santo Andre, SP, Brazil*
*E-mail: renato.kinouchi@ufabc.edu.br*

## SHORT ABSTRACT

In this work we reevaluate and elaborate Crick-Mitchison's proposal that REM-sleep corresponds to a self-organized process for unlearning attractors in neural networks. This reformulation is made at the face of recent findings concerning the intense activation of the amygdalar complex during REM-sleep, the involvement of endocannabinoids in synaptic weakening and new neural network models with itinerant associative dynamics. We distinguish between a neurological REM-sleep function and a related evolutionary/behavioral dreaming function. At the neurological level, we propose that REM-sleep regulates excessive plasticity and weakens over stable brain activation patterns, especially in the amygdala, hippocampus and motor systems. At the behavioral level, we propose that dream narrative evolved as exploratory behavior made in a virtual environment promoting "emotional (un)learning", that is, habituation of emotional responses, anxiety and fear. Several predictions of the unlearning idea that are at variance with the memory consolidation hypothesis are discussed.

## LONG ABSTRACT

In this work we reevaluate and elaborate Crick-Mitchison's proposal that REM-sleep corresponds to a self-organized process for unlearning attractors in neural networks. This reformulation is made at the face of recent findings concerning the intense activation of the amygdalar complex during REM-sleep, the involvement of endocannabinoids in synaptic weakening and new neural network models with itinerating associative dynamics. We distinguish between a neurological REM-sleep function and a related evolutionary/behavioral dreaming function. At the neurological level, we propose that REM-sleep regulates excessive plasticity and weakens over stable brain activation patterns, especially in the amygdala, hippocampus and motor systems by over activating the cannabinoid system. This proposal is motivated by the high density of endocannabinoid receptor CB1 in regions highly active during REM, the high level and efficiency of endocannabinoids during sleep and unlearning-like effects of CB1 receptor activation. At the behavioral level, we propose that dream narrative evolved as exploratory behavior made in a virtual environment promoting "emotional (un)learning", that is, optimization of emotional responses, fear learning and anxiety level. Several predictions of the unlearning idea that are at variance with the memory consolidation hypothesis are discussed. Our proposal is illustrated by an itinerant associative memory model (a Hopfield model with transition between attractors and self-organized unlearning dynamics) that mimics William James' dynamical ideas about the *stream of thought*.

*Taking a purely naturalistic view of the matter, it seems reasonable to suppose that, unless consciousness served some useful purpose, it would not have been superadded to life. Assuming hypothetically that this is so, there results an important problem for psycho-physicists to find out, namely, how consciousness helps an animal, how much complication of machinery may be saved in the nervous centres, for instance, if consciousness accompany their action* (William James).

*In dreaming processes man exercises himself for his future life* (Frederich Nietzsche).

## I. Introduction

One century after Freud's *The Interpretation of Dreams* (Freud 1899) neuropsychologists cannot yet answer if dreams should be interpreted. Although probably there are as many theories about dream function as dream researchers, recent advances in neuroimaging, neurobiology and computational neuroscience have contributed to a more focused study of dreams and their main neurological correlate — Rapid Eye Movement (REM) sleep. Here we follow Hobson and McCarley (1977) and define dreams as *a mental experience, occurring in sleep, which is characterized by hallucinoid imagery, predominantly visual and often vivid; by bizarre elements due to such spatiotemporal distortions as condensation, discontinuity, and*

*acceleration; and by a delusional acceptance of these phenomena as "real" at the time that they occur.* We assume the conventional correlation between REM-sleep and strong dreaming experience (Hobson et al. 2000), although a strict correspondence is not crucial for our purposes.

There are two kinds of questions relative to REM-sleep and dreams. The first are neurophysiologic questions: what are the mechanisms related to REM-sleep generation, what brain regions are activated or inhibited, what neurotransmitters are involved, what mechanism produces motor paralysis during REM-sleep etc. Today, we have a large, although incomplete, body of knowledge of this kind (McCarley 1998; Hobson et al. 2000; Stickgold et al. 2001). The second type refers to functional (or perhaps evolutionary) questions. Why do mammals and birds have REM-sleep but other animals do not? Why so much REM-sleep activity occurs in the fetus and newborn? Are the functions of REM-sleep and dreams the same? What is the function of dreams after all?

We can invalidate this last question by making the skeptical proposal that dreams have no function. After all, although cardiac contractions have the function of blood pumping, cardiac sounds probably do not have any functional meaning. So, dreams could be the brain analogue of cardiac sounds, an epiphenomenon of brain periodic activation during sleep. Notwithstanding this possibility, the search for some functional role not only for REM-sleep but also for dream narrative continues. William James's evolutionary argument that consciousness should have some function is not logically conclusive but continues to be "reasonable" if one substitutes "consciousness" by "dreams = consciousness during sleep".

The activation-synthesis model of Hobson and McCarley (1977) and the activation-input-modulation model of Hobson et al. (2000) are examples of mostly descriptive, not functional, accounts. These authors emphasized the endogenous activation of the forebrain by the Pons-Geniculate-Occipital (PGO) waves during REM-sleep and correlated the activation of specific brain regions (visual cortex, motor system, vestibular system etc.) to typical dream contents. They also proposed that the dream narrative emerges as a tentative and partial synthesis, elaborated in higher associative centers, of such stormy activation. Although Hobson (Hobson et al. 2000) and McCarley (1998) give a detailed account of REM-sleep mechanisms, especially the possible relevance of aminergic/adrenergic balance, and do a wonderful job by correlating brain states with dream phenomenology, they have little to offer for dream/REM function, suggesting some kind of periodic reactivation necessary to the maintenance of poorly used memories and neural circuits (McCarley 1998), or some kind of reprocessing of novel cortical associations (Hobson et al. 2000).

Jouvet advanced the hypothesis that REM-sleep has developmental functions related to the imprinting of genetic behavioral programs into the neural architecture (Jouvet 1999). In a similar way, Changeaux and Danchin (1976) suggested a relation between REM-sleep and synaptic stabilization during development. In both hypotheses, adult dreams are possibly non-functional vestiges from developmental mechanisms (Changeaux 1985) although a similar function to be performed on the mature brain is conceivable (Jouvet 1999). We call this set of ideas as the *Developmental Hypothesis*.

The idea that REM-sleep and dreams have some relation to memory and learning is somewhat old but received renewed attention in the past decade. Winson (1990; 1993) called attention for the appearance of hippocampal theta rhythm in rodents during REM-sleep. Since theta waves are also present in the rodent exploratory behavior, he suggested that information gathered during spatial exploration is reprocessed during dreams. Louie & Wilson (2001), McNaughton (2000) and others also found evidence of replay, during sleep, of hippocampal ensembles that were active during diurnal exploratory behavior.

McClelland (1998) also emphasized the role of hippocampal activation during REM-sleep. In his view, hippocampus serves as a buffer and at same time as a filter for recent learning experiences. New memories are transferred to the cortex during sleep, interleaved with old ones, in order to prevent the catastrophic interference that could occur if strong recent experiences are directly imprinted into cortical networks. This kind of idea is known as the *Memory Consolidation Hypothesis* (MCH) for dream function and has been largely promoted in the literature; see reviews (Stickgold et al. 2001; Maquet 2001). However, although MCH inspired innumerous studies, the evidence favoring it is weak and contradictory, as reviewed by Vertes and Eastman (2000) and Siegel (2001).

A different idea, which we want to explore in this work, is the *Unlearning Hypothesis* (UH) of Crick and Mitchison (1983; 1986; 1995). In 1983, these researchers, at the same time that Hopfield and co-workers (1983), proposed that neural networks are inevitably plagued with parasitic attractors (states produced by mixture and interference between memories) and over-stable ("obsessive") attractors. A possible way for detecting and purging these modes, in a self-organized way without appealing to some supervisor mechanism, is randomly exciting these networks, what could be done, say, by the PGO waves. The following relaxation process leads to the most common or the most stable attractors (because these two kinds of states have larger attractor basins). Then, they postulate that some kind of anti-Hebbian mechanism ("unlearning" or "reversed learning") weakens or eliminates the parasitic states so detected.

Notice that, in the Crick-Mitchison hypothesis, the unlearning idea is central because unlearning promotes a self-limiting (that is, negative feedback) process. This is not symmetrical to re-learning, which produces a positive feedback: if a memory is strengthened, the probability of it being select anew from random stimulation grows, leading to the collapse of the system into a super attractor state. The Crick-Mitchison idea is intriguing because the postulated anti-Hebbian mechanism makes it very distinct from other hypothesis. The nature of such anti-Hebbian process, however, was controversial at the time of the proposal and has lead to some skepticism toward these ideas. We also notice that Crick and Mitchison emphasized (in our view, incorrectly, if we examine the computational model that inspired them) that most dreams are nonsensical garbage, underestimating the narrative coherence and continuity of dreams (for example, repetitive dreams), and also their strong emotional meaning. This occurred because they paid attention mostly to the elimination of spurious associations in the cortex instead of considering with care the effect of unlearning obsessive, over learned or over stable states in the limbic system.

We could not finish this introductory review without commenting Freud's ideas. In a simplified form, Freud's hypothesis is that "Dreams are (disguised) realizations of (repressed) wishes whose function is to diminish the internal wishes pressure during sleep (Freud 1899). The disguised character may appear because some censorship mechanism tries to avoid that those wishes which are unacceptable to the individual emerge in the waking state. Freud recognizes, however, that afflictive dreams full of anxiety seem to be strong counter evidence to his hypothesis: they do not seem to be realizing some wish or desire. Then, Freud needs to appeal to *ad hoc* arguments in order to explain away the presence of fear and anxiety in dreams. For example, he conjectures that some of his patients have afflictive dreams ("counter-desire dreams") because they have a masochist personality or are motivated by the occult "desire" to prove that Freud's theory is wrong! (Freud 1899). However, modern statistical studies of dream content have shown that anxiety indeed is the main affect present in dreams (McCarley 1998; Hobson et al. 2000; Stickgold et al. 2001). It is also very interesting to notice how much Freud's dreams related in (Freud 1899) are plenty of anxiety. Freud's "counter-desire" dreams, afflictive dreams, recurrent nightmares, anxiety dreams about future events etc. are a central concern in our theory because they are prominent dreaming experiences and challenge both Freud's ideas and the MCH.

Neuroimaging data taken during REM-sleep shows that the amygdalar complex, the hippocampus and the basal ganglia are the most activated brain regions during REM sleep (Maquet et al. 1996; Braun et al. 1997; Hobson et al. 2000). It is known that the amygdala has important relation to the formation and activation of emotional memories and anxiety/ruminant thoughts (Rogan et al. 1997; Armony et al. 1997; LeDoux 1998). On the other hand, hippocampal activation is commonly related to exploratory behavior and spatial memory (Louie & Wilson 2001; McNaughton 2000).

Recently, the cannabinoid system has been implicated in short-term synaptic weakening (Wilson et al. 2001; Sullivan 2000), inhibition of electrical synapses (Boger et al. 1998), blocking of synapse stabilization (Kim & Thayer 2001), blocking of long term memory formation (Stella 1997) control of stress response (Martin et al. 2002), fear learning extinction (Marsicano et al. 2002) and other apparently unrelated functions (locomotor control, pain suppression, immunological responses). There is also evidence that connects endocannabinoids to sleep induction and REM-sleep control (Murillo-Rodriguez et al. 1998; Murillo-Rodriguez et al. 2001). These functions could be linked to the recovery from aversive outcomes in exploratory behavior.

Our aim is to put all these new data in a framework inspired by Crick-Mitchison unlearning hypothesis and new itinerant memory models. Motivated by these new findings, we propose that:

1. REM-sleep promotes weakening of over stable attractors in neural networks prone to excessive plasticity such as the amygdala, hippocampus, cerebellum, basal ganglia etc. It is a general homeostatic mechanism to prevent excessively recurrent neural states and actuates even in plastic neural tissues not related to conscious information processing (such as the habit forming locomotor system). This plasticity control is done by a synaptic weakening mechanism activated during REM-sleep and works even in the absence of phenomenal dreams.
2. Dream narrative generation is a particular aspect of such plasticity control. Its evolutionary and developmental emergence is posterior to the development of the REM synaptic weakening mechanism. Dream generation may be implemented by simple activation and associative itinerancy in networks related to exploratory behavior.
3. Dreams are simulations of exploratory behavior whose main function is to elicit emotional responses and promote their habituation. It is immaterial if the dream content refers to true memories or not. The important point is that the dream content is able to elicit emotional responses (fear, anxiety, obsessive wishes or thoughts etc.) in the amygdalar complex to be subjected to the habituation process.

4. Synaptic weakening is done by strong activation of the endocannabinoid system during REM. Dream amnesia reflects this activation.
5. Synaptic weakening made during REM-sleep constitutes an essential step in developmental synaptic pruning, not synaptic stabilization.

We must recognize that our ideas are not incompatible with Freud's ones, since wishes are brain states potentially obsessive. The realization of a wish through a virtual simulation may, as proposed by Freud, may diminish its obsessive power (a habituation effect). Wishes also may elicit some anxiety: in several languages, to be anxious refers both to positive (to desire) and negative (to fear) affects. Endocannabinoids have ansiolitic properties and also can activate (indirectly) the reward system (Martin et al. 2002), inducing a simulated satisfaction. Also the idea of repression of dream content due to its anxiogenic potential is somewhat akin to the idea of unlearning: trying to make those states repellors instead of attractors.

Our theory, however, intends to be more general: there are several scenarios, not related to inner desires or wishes but linked to mammalian exploratory behavior, that elicit anxiety and excessive brain plasticity. For example, exploration of (present) unknown environments, evocation of (past) traumatic memories or rumination about (future) possible aversive situations. In our theory, these scenarios would be fundamental dream themes.

The presumed dream function is also more general: not simply to discharge stress by realizing, even in a disguised way, some wishes, but to habituate fear, anxiety and obsessive states (which sometimes accompany wishes but in general have other origins). In a nutshell, we propose that dream narrative function is akin to desensitization therapy (Marks & Tobena 1990), done in a simulated and virtual environment (the dream). Since dream narratives are simulations, they expose the subject to charged emotional experiences in a secure way. This has evolutionary advantages over real experience or even mammalian play behavior. This last point has also been emphasized by Revonsuo evolutionary dream theory (Revonsuo 2000), although we stress here emotional (un)learning (extinction of emotional responses) instead of cognitive learning of skills in virtual environments.

The paper is organized as follow. In Sec. II we present our reformulation of the Crick-Mitchison unlearning hypothesis. The hypothesis is connected to recent experimental findings, put in a broader context and compared to alternative dream theories. In Sec. III, we introduce a simple computational model of self-organized unlearning which includes dream narrative as an itinerant associative dynamics. In Sec. IV we discuss the results obtained with the computational model. As a spin-off, we get an intriguing suggestion about replay and acceleration in dream phenomenology. In Sec. V we discuss a possible biomolecular foundation for the unlearning process: a strong activation of the cannabinoid system during REM-sleep. In Sec. VI, we present some predictions derived from the Unlearning Hypothesis and suggest new experimental tests. The last section contains a summary of our findings.

## II. Reformulating the Crick-Mitchison hypothesis

Since its original formulation, there occurred at least three developments relevant to Crick-Mitchison's unlearning hypothesis: 1) the application of brain imaging techniques to the study of REM-sleep (Maquet et al. 1996; Braun et al. 1997); 2) advances in the understanding of the neurobiology of extinction, habituation, synaptic weakening and other unlearning-like phenomena (Marsicano et al. 2002); 3) a deeper analysis of unlearning algorithms in attractor neural networks (vanHemmen 1997). These algorithms were the basis for the ideas developed by Crick and Mitchison.

We claim that these developments solve several weak points of the original Crick-Mitchison hypothesis. These deficiencies are related to: a) what is unlearned; b) why it is unlearned; c) how it is unlearned. This is fully elaborated in the next subsections. Comparison with the memory consolidation hypothesis (MCH), the main rival of the unlearning hypothesis (UH), is made in parallel.

## A. What is unlearned?

In the initial formulation of unlearning algorithms, the basic model was an attractor network with a simple auto-associative synaptic matrix, the Hopfield model (Hopfield et al. 1983) This network can store up to *0.14 N* patterns (representing memories or other meaningful brain states), where $N$ is the number of neurons. Above this threshold, there occurs a load catastrophe: the memorized patterns turn out unstable and a myriad of mixed (combination of three or more patterns) and spurious states ("spin glass" states that have no correlation with the learned patterns) appear (Krogh et al. 1991).

Hopfield and co-workers noticed that if we randomly excite the network and let it relax, mixtures and spurious states are preferably found. Then, if we add to the synaptic matrix an anti-Hebbian term, after several cycles of this unlearning procedure, we recover positive stability for the desired patterns, even well above the *0.14 N* load.

When Crick and Mitchison published their dream hypothesis (with Hopfield and co-workers publishing the unlearning algorithm in the same *Nature* issue), they naturally identified the dream content to the mixture and spurious states. This means that dreams were initially thought as a succession of uncorrelated and meaningless states. This initial formulation emphasized the bizarreness of dream content, the condensation (mixture states) phenomenon and downgraded the apparent coherence of dream narratives. A common formulation at that time was "dreams are garbage which contains no information and need to be eliminated instead of remembered", leading to understandable resistance by dream researchers and psychotherapists.

Later, it has been recognized that mixture and spurious states are not of primary importance in attractor networks, since, by using incremental learning algorithms instead of the simple correlational matrix of the Hopfield model, one can store up to the theoretical limit of *2N* patterns without the load catastrophe (Krogh et al. 1991; Bouten et al. 1995). However, studies about the unlearning algorithm continued, mainly due to the interesting self-organized character of the unlearning procedure and its biological flavor (vanHemmen 1997).

We think, however, that from a biological point of view, the problematic states in attractor networks are not spurious or mixture states but patterns over imprinted due to strong synaptic modification. This problem has been addressed by some authors, even in the first papers about unlearning algorithms (Clark et al. 1984). These over learned states are, in a sense, obsessive: the recovery flux is monopolized by them and other normal patterns have their stability and attraction basins diminished.

In biological networks, the origin of these over learned states is varied: fear learning produces strong conditioned associations and emotional responses; traumatic events may lead to strong emotional episodic memories common in Post-Traumatic Stress Disorder (PTSD); ruminative (recurrent) thoughts are common in depression and anxiety disorders; sexual rumination certainly may turn out obsessive. Highly pleasurable experiences could also induce strong conditioned responses and addicting behavior. Obsessive-compulsive disorder involves recurrent activity in motor systems. Strong learning protocols (such as intensive training in spatio-visual games like chess, Tetris and videogames) induce spontaneous recurrent hypnagogic imagery (Stickgold et al. 2000). Repetitive motor activation also may induce excessive imprinting in motor areas. On the overall, brain plasticity seems to lead easily to obsessive dynamics, from the common experience of catchy jingles and intruding music to stereotyped motor activation (tremor, tics, Tourete syndrome etc.) and obsessive-compulsive thought and behavior. And, as it is well known, the most plastic brain tissues are the amygdala, the hippocampus and the cerebellum. It seems that a prime problem for these regions is not how to learn but how to not over learn.

But if REM-sleep/dreams refer to an unlearning procedure that basically try to manage over imprinted patterns then dreams mainly shall be composed by meaningful, emotional content. For example, if dreams correspond to the brain tentative of extinction of fear learning, then the usual presence of anxiety and threatening scenarios in dreams and the repetitive nature of nightmares (Hartman 1998) make sense. Dream content will not be formed by spurious or mixture states, but will be related to obsessive or potentially obsessive states which need to be detected and weakened. Dreams will be "royal roads" to emotionally charged memories, fears and obsessions stored in the amygdala. **What must be unlearned is not the dream content, but the excessive emotional response to the dream content.**

## B. Why is it unlearned?

We have argued that an important problem in attractor networks is that over imprinted states (which we call "strong patterns" from now) have very large and deep attractor basins. They not only monopolize the recovery dynamics (a large set of initial conditions activates the strong patterns) but, indirectly, the stability of other patterns is reduced. We think that theses states are the primary targets for unlearning during REM-sleep. In Crick and Mitchison's words: "We dream to forget". Our hypothesis is slightly more specific: "We dream to weaken over stable neural states". We dream as a part of an endogenous homeostatic mechanism that prevents excessive plasticity and obsession.

We remark that, in our view, unlearning is needed not only for fear conditioned responses but also for over imprinted episodic memories, over trained procedural behavior, pleasurable behavior, sexual ruminations etc, in the measure that they also might turn out obsessive: lovers dream with each other in order to habituate love obsession! However, since fear learning is the most studied model for the formation of strong memories/associations, and recent findings about the role of endocannabinoids in fear learning extinction (Marsicano et al. 2002) enable us to make experimental predictions, we concentrate from now in dreams involving fear and anxiety (which indeed are the main affects present in dreams (McCarley 1998)).

But is REM-sleep only part of a homeostatic mechanism for the control of excessive plasticity? We believe that this is the primary function of REM-sleep (which is present even in low mammals, fetuses and newborns where dream narrative is absent). But biological features frequently are selected because they implement (or contribute to) several functions, not all operating at the same time: for example, play behavior presumably has been selected in mammals because it contributes both for cognitive learning (how to map and control the environment), social learning (how to interact with others) and emotional learning (how to control emotional responses). The function of play may also be different for small children, mature children and adults. We believe in a similar role for dream narrative: like play behavior, dreaming at more mature stages provides a secure mechanism for emotional (un)learning, that is, habituation and extinction of excessive emotional response to aversive situations. REM-sleep supports this function by furnishing the basic unlearning mechanism.

We notice that fear and anxiety are main affects during exploratory behavior. So, we conceive dreams not only as having a defensive, self-repairing role, preventing obsession from excessive plasticity (say, traumatic memories). Dream narrative seems to be an exploratory behavior made off-line during sleep, where the organism not only explores virtual environments constructed from the past but also from its present and future concerns (anxiety thoughts and fears). Anxiety dreams are mainly simulations of threatening scenarios, already encountered in waking life or not. For an elaborated defense of this hypothesis, see Revonsuo (2000).

This means that, like juvenile play and other simulated behavior, dreams are future-oriented or prospective: habituation of excessive emotional response to aversive stimuli prepares the organism to future encounters with more aversive situations. This contrasts to the past-oriented character assumed by the memory consolidation hypothesis but accords with the frequent case of dreaming about future-related events. Dreams do not consolidate the past but explore the future by simulating aversive scenarios: a concrete realization of Nietzsche quoted idea about dream function.

## C. How is it unlearned?

For unlearning strong patterns we must detect them and change the part of the synaptic matrix responsible for their stability. These strong patterns must be activated before weakening them if we desire selective unlearning instead of simple passive decay. But their replay should not collaborate to their memory consolidation. These contradictory requirements may be the reason for dreams to occur under inhibition of mechanisms for long term memory formation and attention processes.

We think that the detection of strong patterns is done automatically under the dream associative process. In the computational model below, we assume that dreaming is an uncontrolled Jamesian stream of thought (James 1892), a staccato flow of memories, thoughts and mental images in Walter Freeman dynamical description (Freeman 2000). This flow is not purely random but constitutes a narrative (or several narratives) full of associative transitions and emotional content. This is realized in our model by a continuous but chaotic exploration of network states made autonomously due to short-term activation-dependent synaptic weakening.

This continuous wandering is affected by the relative stability of the embedded patterns. Over imprinted patterns finally dominate the dynamics, that is, initial random activation of the network leads, after a transient, to preferential wandering over these states (memories or associations). We think that this chaotic itinerant dynamics gives a basis for modeling the dreaming state as a (noisy) associative narrative that walks toward an emotional climax.

The strong patterns are then weakened by the unlearning process. This unlearning is not a passive synaptic decay, but is an active process, selective and associative, in the sense that only states visited during the dream induce anti-Hebbian synaptic weakening. The strong patterns must be replayed, elicited anew for anti-Hebbian processes to work. This view is compatible with the recent findings about the importance of replaying the aversive cues for endocannabinoid production and concomitant extinction of fear learning (Marsicano et al. 2002). The possible relationship between anti-Hebbian unlearning and endocannabinoids will be fully elaborated in Sec. V.

We make a remark about an experimental cornerstone of the Memory Consolidation Hypothesis: the replay of hippocampal place cells in rats (Louie & Wilson 2001; McNaughton 2000). Under the Unlearning Hypothesis, this replay does not mean memory replay for consolidation, but corresponds to the elicitation of environmental cues related to a simulated aversive situation. In our view, this replay of place cells only sets the scenario, like a human dreaming of being persecuted by a strange inside his familiar house.

## D. Comparison with the developmental hypothesis

In the fetus and newborn, the memory systems and synapses are very plastic. As already observed by Crick and Mitchison, probably most of the initial network activation induces accidentally over imprinted states and spurious associations which should be unlearned. We suggest here a stronger connection between REM-sleep and development: unlearning may be a necessary step toward the developmental massive synaptic pruning that sculpts the mature brain (Rakic et al. 1994; Innocenti 1995; Barbato & Kinouchi 2000).

So, according to our view, REM-sleep in fetuses and immature children detects over imprinted states for weakening them. It performs the primary role (control of over stable patterns) but not the secondary role (emotional (un)learning, made by dream narrative which probably is absent in small children). Contrasting to the developmental hypothesis, we propose that extensive REM-sleep does not promote synaptic stabilization or imprinting of genetic memories, but synaptic weakening, which is known to be a necessary step for synaptic pruning and axon withdraw. This could be tested by searching for temporal correlation between synaptic and REM-sleep density curves. Anew, we have synaptic anti-Hebbian weakening before pruning, not memory consolidation.

## D. Comparison with the Memory Consolidation Hypothesis

Spatial exploratory behavior correlates well with quantitative analysis of dream content (Revonsuo 2000). Anxiety and fear is a principal component in exploratory behavior, both in a past and in a future sense: bad outcomes of past exploratory behavior and the expectation of bad outcomes in new environments are the stuff of fear and anxiety. The statistics of dream content shows a high proportion of "walking dreams" in new, possibly threatening environments. There is also a higher proportion of prey behavior ("chase dreams") that elicit the classic freezing, running and aggressive defense behaviors. These themes are not simple memories to be consolidated or Freudian (masochistic) desires, but threatening scenarios that make sense from an evolutionary point of view (Revonsuo 2000).

Revonsuo "virtual exploratory behavior" hypothesis explains why the specific areas listed by Hobson and McCarley (Hobson & McCarley 1977; McCarley 1998; Hobson et al. 2000; Stickgold et al. 2001) are the most activated during REM-sleep: brain areas related to walking function (motor cortex, basal ganglia, cerebellum, vestibular area (Braun et al. 1997)), visual areas in primates (but with inactivation of primary V1 area and frontal areas, forming a visual/limbic closed system (Braun et al. 1998)), olfactory bulb in other mammals (Louie & Wilson 2001). Basically, they are related to exploratory behavior and are co-activated with emotional response in amygdala and spatial memory in hippocampus (Maquet et al. 1996; Braun et al., 1997). We also remember that, in the waking animal, theta waves are related to exploratory behavior and PGO waves are elicited in startling responses.

Contrasting to MCH, we conceive the dream as not being a simple rehearsal of previous experience. The simulated exploratory behavior is "new" in the sense that the threatening scenarios and the corresponding emotional responses, although stereotyped, are not simple replays. They are better conceived as emerging from the primary excitation of the amygdala: this excitation leads to an associative walk between the amygdalar strong attractors.

We think that subjects with PTSD have vivid dreams with the original trauma trying to habituate their emotional response (Rothbaum & Mellman 2001). The relative failure of this homeostatic mechanism may be related to a low production of cannabinoid receptors (Marsicano et al. 2002). After all, traumatic memories need not to be consolidated, but weakened. We also observe that dreams of subjects with PTSD not only replay the primary experience but frequently refer to threats in the present, contain distorted features and drift away from the original theme as recovery progress (Esposito et al. 1999; Hartmann 1998). This temporal evolution is fully compatible with the emotional response Unlearning Hypothesis. It is also curious that virtual reality simulations are being used for treatment of PTSD (Rothbaum et al. 2001) and that a treatment that mimic REM-sleep (EMDR – Eye Movement Desensitization and Reprocessing) is perhaps the most efficient psychotherapy to PTSD (Stickgold 2002).

We emphasize that dreaming with future oriented events also stays in clear contradiction to the Memory Consolidation Hypothesis. As further examples of future oriented dreaming that explores new situations, we may cite some common dream themes: examination dreams and marriage dreams. These dreams mainly occur before the event date (for example, to dream with a PhD thesis examination before its occurrence) and usually are full of anxiety (usually the dream involves vexing outcomes). Another future oriented theme is a pregnant woman dreaming with an ill-formed baby or a mother dreaming with her children being in danger: no true memories (nor masochist desires) to be consolidated here.

## III. Computational Model

Crick and Mitchison developed their ideas from suggestions given by a simple computational model (Hopfield network). Let us concretize our ideas by using an improved neural network model. We prefer to call these models as "qualitative computational models", in a sense similar to "qualitative analysis of differential equations". We want to examine generic and robust dynamical behaviors, not quantitative details.

For those not familiar with models as used in theoretical science, we clarify the methodological function of (qualitative) computational models. They do not intend to prove (or disprove) a hypothesis, to substitute experiments, or even to simulate biophysical mechanisms (as done in computational neuroscience). Qualitative models intend to realize a verbal model in a more concrete way, to demonstrate that the verbal model "works", showing that its assumptions are indeed sufficient for producing the desired behavior.

Implementing a computational model often aid to make explicit some tacit assumptions incrusted in verbal models. It also adds plausibility to the original idea because it may show that the behavior of interest is robust if model parameters are changed. Another fruitful outcome is the appearance of unexpected phenomena: behaviors not expected (or even counter-intuitive) given the verbal model may turn out prominent in the computational model. These unexpected behaviors are frequently produced by feedback effects. As is well known, verbal models usually have difficulty to describe circular causality so common in dynamical systems.

## A. Itinerant associative dynamics: implementing William James's "Stream of Thought"

One of the cornerstones in the resurgence of neural networks models in the 80's is the content addressable memory (CAM) model of Hopfield (1982; 1983; 1984). This model popularized a paradigm where memories and meaningful brain states are represented as state vectors that are (fixed point) attractors: the relevant states are stationary spatial patterns of neural firing in the network. Later, extensions of this kind of attractor network have been proposed that implement transitions between patterns, the so called itinerant memory models (Horn & Usher 1991; Adachi & Aihara 1997; Adachi & Aihara 1999). In itinerant models, there are no more true fixed point attractors: a slow dynamics, usually implemented as neuronal fatigue (or, as in our model, as synaptic regulation), induces transitions between the former attractors.

From a historical perspective, this kind of itinerant dynamics indeed may be viewed as a computational realization of original ideas from William James more than a century ago. A strong emphasis in James writings is his recurrent comparison between the complexity of the thought flux and the analogous complexity of a turbulent fluid. The following examples are typical of James writings (James 1884; 1892):

*What must be admitted is that the definite images of traditional psychology form but the very smallest part of our minds as they actually live. The traditional psychology talks like one who should say a river consists of noting but pailful, spoonful, quartpotsful, barrelsful, and other molded forms of water. Even were the pail and the pots all actually standing in the stream, still between them the free water would continue to flow. It is just this free water of consciousness that psychologists resolutely overlook. Every definite image in the mind is steeped and dyed in the free water that flows round it.*

James analogy of the "stream of thought" as a complex fluid is not an old fashioned idea but a very fortunate one, from a physicist point of view. Indeed, although the local laws for fluid movement are well known, their strong non-linearity precludes a global understanding of the complex behavior of fluids. Pattern formation such as stable vortices and other structures are ill understood, and the origin of complex patterns and turbulence in fluids is one of the greatest challenges to present day physics.     By exposing his analogy of mental processes with complex physical systems, James wants to recover the richness of the thought process from rationalist models based on a manipulation of atomic, isolated, timeless symbols, a perspective dominant in XIX century which yet today informs a large part of cognitive science:

*The demand for atoms of feeling, which shall be real units, seems a sheer vagary, an illegitimate metaphor. (…) There is no reason to suppose that the same feeling ever does or can recur again. The same thing may recur and be known in an indefinite number of successive feelings; but does the least proof exist that in any two of them it is represented in an identical subjective state? All analogy points to the other way. For when the identical thing recurs, it is always thought of in a fresh manner, seen under a somewhat different angle, apprehended in different relations from those in which it last appeared. (…) However it may be of the stream of real life, of the mental river the saying of Herakleitos is probably literally true: we never bathe twice in the same water there. How could we, when the structure of our brain itself is continually growing different under the pressure of experience? For an identical feeling to recur, it would have to recur in an unmodified brain, which is an impossibility. The organ, after intervening states, cannot react as it did before they came.*

In other words, for James, memories, feelings and thoughts are complex spatio-temporal patterns which cannot be isolated from an underlying neural activity as vortices cannot be isolated from the river stream which gives origin to them. His emphasis on this distributed and extended nature of mental states (a high dimensional activation pattern), on the bounded but non-recurring activity (a chaotic attractor behavior) and long-range temporal correlations typical of complex systems is surprisingly contemporaneous. James contrasts psychologies based in grammatical-like symbol manipulation with ideas inspired by the complex behavior of extended dynamical systems, favoring the last description. It seems that the cognitivist versus dynamicist controversy (van Gelder 1998; Sternad 2000; Kinouchi 2001) has his roots at least in William James times.

Of particular interest to our modeling effort is James description of the thought flux as being at the same time continuous but not homogeneous, a sequence of transitions between metastable states:

*When we take a rapid general view of the wonderful stream of our consciousness, what strikes us first it the different pace of its different portions. Our mental life, like a bird's life, seems to be made of an alternation of flights and perchings (...). The resting places are usually occupied by sensorial imaginations of some sort, whose peculiarity is that they can be held before the mind for an indefinite time, and contemplated without changing; the places of flight are filled with thoughts of relations, static or dynamic, that for the most part obtain between the matters contemplated in the periods of comparative rest.*

*Let us call the resting-places the "substantive parts" and the places of flight the "transitive part", of the stream of thought. We may then say that the main end of our thinking is at all times the attainment of some other "substantive" part than the one from which we have just been dislodged. And we may say that the main use of the transitive parts is to lead us from one substantive conclusion to another.*

In our model, the substantive states are the embedded patterns usual in attractor networks. The transition between substantive states will be induced by a short-term anti-Hebbian term although other neuronal fatigue mechanisms could be used.

Another important point emphasized by James is the fuzzy and mixed character of each actual mental representation, which shall be realized in our model by the fact that the neural network state at a given time $t$ may have simultaneous correlation to several embedded memory vectors:

*If the image comes unfringed it reveals but a simple quality, thing or event; if it come fringed it reveals something expressly taken universally or in a scheme of relations. The difference between thought and feeling thus reduces itself, in the last subjective analysis, to the presence or absence of "fringe". And this in turn reduces itself, with much probability, in the last physiological analysis, to the absence or presence of sub-excitements of an effective degree of strength in other convolutions of the brain than those whose discharges underlie the more definite nucleus, the substantive ingredient, of the thought — in this instance, the word or image it may happen to arouse. (...) It is in short the re-instatement of the vague to its proper place in our mental life which I am so anxious to press on the reader's attention.*

James also tries to link his ideas to the physical models of brain activity of his time:

*The best symbol for it seems to be an electric conductor, the amount of whose charge at one point is a function of the total charge elsewhere. Some tracts are always waning in tension, some waxing, whilst other actively discharge. The states of tension, however, have a positive influence as the discharges in determining the total condition (...). But as the distribution of brain-tension shifts from one relative state of equilibrium to another, like the aurora borealis or the gyrations of a kaleidoscope, now rapid and now slow, it is likely that the brain faithful psychic concomitant is heavier-footed than itself, that its rate of change is coarser-grained, that it cannot match each one of the organ's irradiations by a shifting inward iridescence of its own? But if it can do this, its inward iridescences must be infinite, for the brain-redistributions are in infinite variety. If so coarse a thing as a telephone-plate can be made to trill for years and never reduplicate its inward condition, how much more this be the case with the infinitely delicate brain?*

Modern advocates of chaotic itinerancy in memory models seem not to be aware of the honorable philosophical and psychological tradition behind their ideas. Since it has been the reading of James that has inspired this work, we would like to amend this somewhat priority injustice by calling our particular itinerant memory model as the "James Machine".

## B. The computational model

The itinerant flux is implemented in the following associative memory network. A set of $N$ units (each unit may represent a neural population instead of a single neuron) are fully connected. Each unit is modeled by a continuous state variable (average firing rate or "activation level") $S_i$ ($i=1,...,N$) which are updated in parallel at discrete time steps as:

$$S_i(t+1) = G\left(\sum_{j=1}^{N} J_{ij}(t)S_j(t) - \theta\right) \qquad . \qquad (1)$$

The transfer function $G(x)$ could be any sigmoid function of the total input $x$ with image in $[-1,1]$ (negative values mean activity bellow some spontaneous average level); $\theta$ is a bias term related to the cell excitability threshold. For computational convenience (absence of calculation of exponentials) we perform our simulations with the function:

$$G(x) = \frac{\gamma x}{1+|\gamma x|} \qquad . \qquad (2)$$

The initial synaptic matrix $J_{ij}(0)$ has the form of a weighted correlation matrix (Clark et al. 1984; Sejnowsky et al. 1989) storing $P$ vector patterns $\boldsymbol{\xi}^{\mu} = \xi_i^{\mu}$, $(i=1,...,N; \mu=1,...,P)$ :

$$J_{ij}(0) = \frac{1}{PN}\sum_{\mu=1}^{P} F^{\mu}\xi_i^{\mu}\xi_j^{\mu}$$

(3)

The patterns $\boldsymbol{\xi}^{\mu}$ can be interpreted as memories, but more generally they represent metastable network states (*James's substantive states*). For simplicity, we choose $\xi_i^{\mu} \in \{1,-1\}$ randomly. In the following we use boldface symbols for vectors and normal symbols for components.

If a direct application to fear learning is desired, we can interpret the pattern $\{ \xi_i \}$ $(i=1,..., N)$ as composed by two sub patterns: $\xi_i^{(S)}$ $(i=1,...,N_S)$ is the internal representation of the aversive stimulus and $\xi_i^{(R)}$ $(i=N_{S+1},...,N)$ is the corresponding emotional response. Since the network is a content addressable memory, partial activation of the fear cues $\boldsymbol{\xi}^{(S)}$ will elicit the activation of the emotional response $\boldsymbol{\xi}^{(R)}$. Notice that we are working in the memory regime $(P < 0.14 N)$, bellow the load catastrophe.

The weight $F^{\mu}$ controls the relative stability of pattern $\boldsymbol{\xi}^{\mu}$. In a biological context, they are induced by factors like attention, surprise, stimulus salience, importance due to aversive context, intensive repetition etc. The initial symmetry condition $J_{ij} = J_{ji}$ is used here for simplicity, but is not important for itinerant memory models and may be relaxed.

The network state $S(t) = \{S_1(t),...,S_N(t)\}$ changes following Eq. (1) in a very fast time scale (milliseconds). As is well known (Hopfield 1982; 1984), a random initial state $S(0)$ relaxes toward some embedded pattern $\boldsymbol{\xi}^{\mu}$. So, in this scenario of strong patterns bellow saturation, it is natural that random initial activation leads to the elicitation of non-random and meaningful states.

The distinctive ingredient of our model is that the synapses $J_{ij}(t)$ are not static but evolve, both in a fast as a slow time scale dynamics, involving anti-Hebbian terms (the biological rationale for anti-Hebbian processes will be discussed latter). The fast time scale dynamics (synaptic regulation) is written as:

$$J_{ij}^{f}(t+1) = (1-1/\tau)J_{ij}^{f}(t) - \eta_f S_i(t)S_j(t)$$

,

(4)

where $\eta_f$ is a small step parameter. The fast dynamics transform the state being visited (usually some embedded pattern $\boldsymbol{\xi}^{\mu}$ into a repellor: the pattern is destabilized. However, this change is not permanent but reversible: the repulsive anti-Hebbian term associated to the visited states is erased by the $(1-1/\tau)$ factor, that is, $\tau$ is a recovery time of the synaptic regulation process.

Without this destabilization dynamics, the system is simply a Hopfield model with graded response units (Hopfield 1984). In that case, if we start from some initial state, the system relaxes fast to some of the stored patterns and remains there. This means that the patterns $\boldsymbol{\xi}^{\mu}$ are stable fixed-point attractors. The synaptic regulation dynamics, however, makes the embedded patterns metastable, that is, the system finally escapes from the present pattern to another one and so on, performing an "itinerant flux". The rapid transient states (which sometimes are combinations of several patterns) between the substantive states correspond to James's *transitive states*.

In the slow time scale, connections evolve as:

$$J_{ij}^s(t+1) = J_{ij}^s(t) - \eta_s S_i(t) S_j(t) \quad ,$$

(5)

Since this process has no recovery time, we have a permanent anti-Hebbian unlearning. The destabilization process must be faster than the unlearning process, because we want that transitions occur before unlearning turns the pattern a permanent repellor. To control plasticity (without forgetting the important attractors) we need pattern weakening, not pattern erasing. This is implemented by setting $\eta_s << \eta_f$.

The fast synaptic component can be initialized from zero $(J_{ij}^f(0)=0)$ and the slow component may start from the initial state given by Eq. (3). Finally, at any time, the total synaptic efficacy is:

$$J_{ij}(t) = J_{ij}^f(t) + J_{ij}^s(t) \, .$$

(6)

Notice that the particular choice of network model (sigmoid units, full connectivity, initial connection matrix, discrete time, parallel update etc.) and even the specific destabilization mechanism are immaterial. The important point is the implementation of transitions between patterns (itinerant dynamics) and long term weakening of visited patterns (to make permanent changes in the synaptic matrix). In more abstract terms, we need a dynamics where patterns are saddle points or, as in our model, fixed points that turn out (reversibly) unstable after being visited, and whose permanent stability is weakened slowly. This could be implemented in other model architectures, for example, in the Hopfield-like model of Sommer and Wennekers (2001) where the basic elements are two-compartmental (Pinsky-Rinzel) biophysical neurons.

## IV. Results and Discussion

### A. Effect of fast synaptic regulation: James itinerant flux

The overall dynamics of the model is the following. With the fast synaptic regulation given by Eq. (4) but without the long-term unlearning process of Eq. (5), the system performs a chaotic itinerant flux between the substantive states $\xi^\mu$. *Chaotic* is used here in the technical sense: not random or disorganized flux, but simply a non-periodic but bounded flux whose long term outcome is sensible to small perturbations.

The weights $F^\mu$ are not permanently modified by the itinerant flux. If there are strong (obsessive) patterns, these attractors dominate the dynamics, creating cycles between them ("meta-basins") and preventing the access to the normal states. For example, we show in Fig.1 the case where three strong patterns *(A,B,C)* with *F=3* dominate over three normal ones with *F=1*. This is detected by following the evolution of the overlaps:

$$m^{\mu}(t) = \frac{1}{N} \sum_{j=1}^{N} \xi_j^{\mu} S_j(t) \qquad , \qquad (7)$$

that measure the correlation between the network state and the $\mu$-th embedded state.

We notice that pattern destabilization favors *associative* itinerancy, that is, transitions between correlated patterns (i. e., that have mutual overlap $M^{\mu\nu} = 1/N \sum_{j=1} \xi_j^{\mu} \xi_j^{\nu}$ of order $O(1)$. We identify this chaotic free associative trajectory with the dream narrative. In our example in Fig.1, patterns *B* and *C* have mutual overlap of *0.1*.

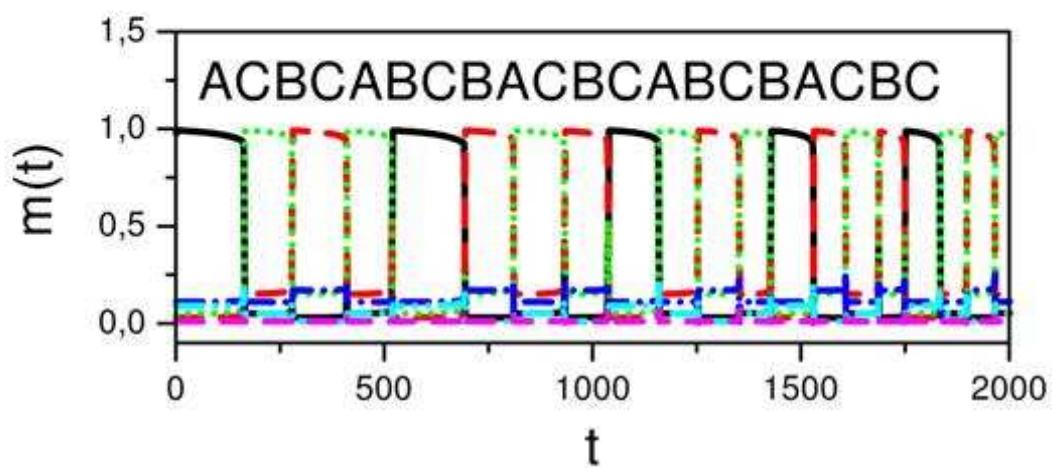

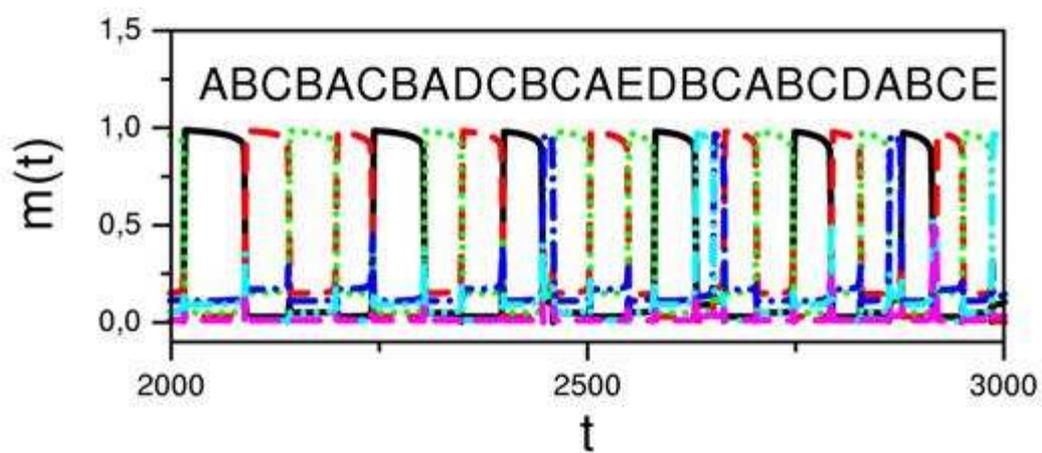

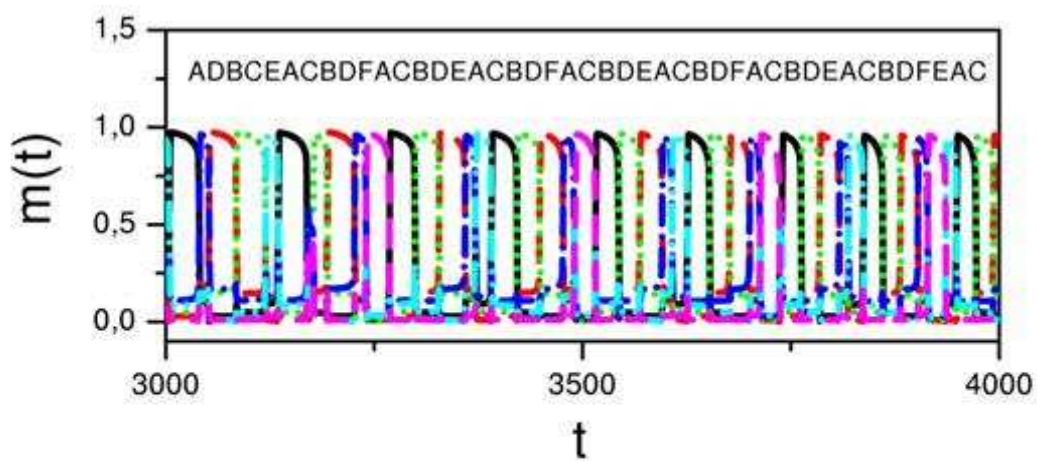

**Figure 1:** Time evolution of the overlaps $m^\mu(t)$ between the network state and the stored patterns during the itinerant dynamics. Parameters: $N = 100$, $\theta = 0$, $\eta_f = 0.02$, $\tau_f = 100$, $\eta_s = 0.001$, three strong patterns {A, B, C} with $F=3$ and three normal patterns {D, E, F} with $F=1$. Notice that we have constructed patterns $B$ and $C$ with a mutual overlap $M^{BC} = 0.1$ and, due to this correlation, the activation of one of them leads to the evocation of the other and vice-versa. The permanent unlearning process starts at $t = 1000$. With progressive unlearning, the transition rate accelerates (notice the different time scales in the panels). Eventually the system escapes from the obsessive meta-basin {A, B, C} and the normal patterns {D, E, F} turn out accessible for the itinerant dynamics. The overall dynamics is chaotic, not cyclic, as can be hinted by the variable order of pattern excitation.

## B. Comparison with adaptive threshold models

Horn and Usher (1991) and Adachi and Aihara (Adachi & Aihara 1997; 1999) developed models of itinerant dynamics where the transitions between patterns is due to the excitability bias term $\theta(t)$. In these models, if the cell is active above a basal level, the bias $\theta(t)$ grows so that the cell accommodates to the input and if the neuron is quiescent, the bias decays as:

$$\theta(t+1) = \theta(t) + \eta\,(S_i(t) - S_R) \qquad , \qquad (8)$$

where $S_R$ is the basal level (say, $S_R = 0.5$).

We have performed simulations with this mechanism (not show). In contrast to our synaptic anti-Hebbian regulation term, the excitability bias mechanism induces not only the desired transitions between different patterns but also a fast oscillation between patterns and anti-patterns. This may be observed also in the Adachi-Aihara simulations (Adachi & Aihara 1999). Since the interesting transitions are between different patterns, the trivial pattern-antipattern oscillation is undesirable and biologically unrealistic. Our dynamics with synaptic regulation does not produce this pattern-antipattern oscillation because both states turn out repellors under the synaptic anti-Hebbian term.

## C. Effect of unlearning: control of the stabilities of attractors

When the slow unlearning dynamics is active, the connection matrix $J_{ij}^S$ slowly changes, making the visited patterns permanently less stable. Since the intensity of weakening is proportional to the time spent in the visited state (which, by its turn, is correlated to the state stability), the patterns with the largest or deep basins are more weakened.

In Fig.1, we show a single dream trajectory (itinerancy plus unlearning) on the previous pathological case of three obsessive states that form a dominant cycle $A$-$B$-$C$. Unlearning starts at step $t=1000$. In Fig.2 we show the effect of unlearning on the permanent synaptic matrix. As a measure of the pattern strength we use the so called "average stability":

$$\Lambda^\mu = \frac{1}{N} \sum_{i=1}^{N} \lambda_i^\mu$$

$$\lambda_i^\mu = \frac{1}{N} \xi_i^\mu \sum_{j=1}^{N} J_j^s \xi_j^\mu \quad , \tag{9}$$

where the $\lambda_i$'s are the local (cell) stabilities. Notice that we measure the time evolution of the patterns stabilities in terms of the slow $J_{ij}^s(t)$ matrix. Changes in the fast synaptic component $J_{ij}^f$ are reversible and do not contribute to permanent unlearning.

Along the time, we see that not only the stability of the obsessive states is diminished but also the stability of the normal states is enhanced (Fig.2). This is an indirect effect due to the fact that all embedded patterns share the same synaptic matrix. As another index of the equilibrium between patterns, we measure the overlap

$$O_H(t) = \sum_{i=1}^{N} \frac{\sum_{j=1}^{N} J_{ij}^s(t) J_{ij}^H}{\left| \mathbf{J}_i^s(t) \right| \left| \mathbf{J}_i^H \right|} \quad , \qquad \left| \mathbf{J}_i \right| = \sqrt{\sum_{j=1}^{N} J_{ij}^2} \quad , \tag{10}$$

between the long term synaptic matrix $J_{ij}^s(t)$ and the simple Hebb matrix $J_{ij}^H = 1/PN \sum_\mu^P \xi_i^\mu \xi_j^\mu$ where all patterns have weight $F^\mu = 1$. The overlap $O_H(t)$ achieves a maximum at the time where the stability of normal patterns is most enhanced (Fig.2, inset).

We also studied the effect of the standard unlearning algorithm: we simply excite the network, perform an unlearning step when the network achieves a fixed point, and repeat the process (Hopfield et al. 1983; vanHemmen 1997). We have found that the itinerant chaotic process leads to more efficient stability equilibration between patterns than this standard algorithm (to be fully reported elsewhere). Exhaustive simulations with an extensive number of embedded patterns should be performed to fully characterize the effect of parameters $\eta_f$, $\eta_s$, $\tau_f$, the strengths $F^\mu$, the fraction of strong patterns etc. (to be reported elsewhere). Notice that a larger number of patterns could be used (proportional to the number $N$ of neurons) so that the itinerancy dynamics could be very rich. Here we used a small (six) number of patterns because we focused only on qualitative and robust behaviors that are sufficient to illustrate our reformulation of Crick-Mitchison hypothesis.

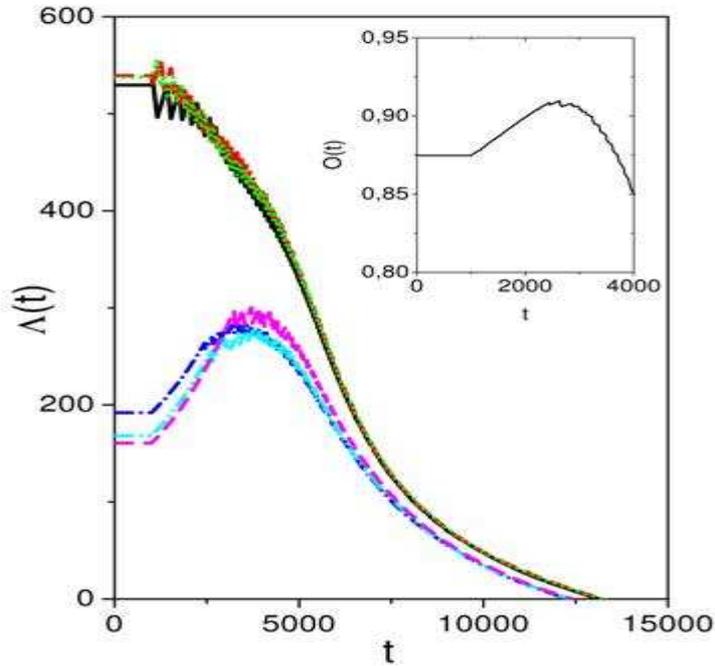

**Figure 2:** Time evolution of the stabilities $\Lambda^\mu$ of the stored patterns during the unlearning process. Notice that the stability of the normal patterns is enhanced when the networks unlearns the strong patterns, achieving a maximum at moderate dream time. **Inset**: time evolution of the normalized overlap $O_H(t)$ between the slow coupling matrix $J_{ij}^s$ and the unbiased ($F^\mu = I$) Hebb matrix.

## VI. A biological candidate for the unlearning mechanism

The weakest point of the unlearning proposal (or perhaps, its strongest prediction) is that if patterns are imprinted in the synaptic matrix by a correlational Hebbian mechanism (Sejnowsky et al. 1989), then an exact (unlearning, reverse learning) anti-Hebbian mechanism is necessary for pattern destabilization (vanHemmen 1997). Up to now, anti-Hebbian mechanisms are not accepted to exist in the brain. However, this is not especially grave since, due to the empiricist pace of neurosciences, even Hebbian mechanisms are not universally accepted as the basis of learning. So, what we need to determine by now is if such anti-Hebbian mechanism gained plausibility after twenty years from its original formulation. We believe that the answer is positive, and propose that the neurobiological basis for anti-Hebbian unlearning is the cannabinoid system.

Contrasting with classical neurotransmission, the cannabinoid system uses retrograde signals. In the central nervous system, it is based on the CB1 receptor (mainly situated at pre-synaptic terminals) and endocannabinoids as N-arachidonoylethanolamine (Anandamide) and 2-arachidonoyl-glycerol (2-AG) that are produced at the post-synaptic terminal (Stella 1997). The main known memory effects of the cannabinoid system are:

a) Short-time weakening of synapses known as Depolarization-induced Suppression of Excitation (DSE) or Inhibition (DSI) (Wilson et al. 2001; Sullivan 2000). In these phenomena, the pre-synaptic fusion of vesicles, which may contain excitatory (AMPA) or inhibitory (GABA) neurotransmitters, is suppressed due to the activation of CB1 receptors in a time scale of tens of milliseconds to seconds. This means that if the post-synaptic neuron is highly firing, it regulates (weakens) its inputs by retro-signaling to the CB1 receptors through endocannabinoids, which are produced on demand due to depolarization of the post-synaptic cell.

b) Blockade of long term potentiation (LTP) or depression (LTD) (Misner & Sullivan 1999; Sullivan, 2000; Stella 1997). This is done both by indirect and direct pathways. It is also been found that mutant mice without CB1 receptors have enhanced spatial memory and LTP in the hippocampus (Bohme et al. 2000), but show persistence phenomena (impaired capacity of unlearn maladaptive behaviors) and increased level of anxiety (Varvel & Lichtman 2000).

c) Cannabinoids are critical ingredients in the extinction of fear learning (Marsicano et al. 2002). Knock-out rats without cannabinoid receptors show increased memory but persistent behavior, slow fear learning extinction rate and high levels of anxiety-like behavior.

d) Cannabinoids inhibit the formation of new synapses (Kim & Thayer 2001).

e) Endocannabinoids also shut down electrical synapses (Murillo-Rodriguez et al. 2001b).

f) Cannabinoids enhance REM-sleep density and impairs memory in rats (Murillo-Rodriguez et al. 1998; 2001).

g) A lipid (Oleamide) that is a strong inductor of sleep and accumulates in sleep-deprived rats also inhibit the hydrolysis of endocannabinoids, promoting higher endocannabinoid efficiency (Murillo-Rodriguez et al. 2001b).

Taken together, these facts put the cannabinoid system as a prime candidate to mediate unlearning during REM-sleep. Endocannabinoids are not stored in vesicles, but are produced on demand by strongly firing of post-synaptic neurons (Stella 1997). We conjecture that, since REM-sleep strongly activates plastic tissues, REM-sleep promotes strong production and liberation of endocannabinoids. The cannabinoid efficiency is further increased by the presence of the sleep-inducing lipid Oleamide. Tissues most activated during REM-sleep also show a high density of cannabinoid receptors: basal ganglia, baso-lateral amygdala (but not central amygdala (Katona et al. 2001)), hippocampus and cerebellum.

The fast synaptic weakening needed for implementing the associative itinerancy that searches the strong attractors (and makes the chaotic dream narrative) could be based in short-time mechanisms like DSE and DSI (Wilson et al. 2001; Sullivan, 2000). Notice that when the pre and post-synaptic cells are depolarized, the synaptic efficiency is diminished by DES, in an anti-Hebbian way. Similarly, if an inhibitory (GABAergic) pre-synaptic cell is silent when the post-synaptic cell is depolarized, DIS also produces anti-Hebbian synaptic weakening (because anti-correlation should induce an increase, not a decrease, of inhibitory efficiency). So, we think that some kind of anti-Hebbian process may be based in the cannabinoid system. However, a detailed account of such anti-Hebbian mechanism is lacking (probably because no experimentalist searched for it).

The fact that endocannabinoids also interfere with long term memory formation (by preventing both LTP and LTD, synapse formation etc.) means that the cannabinoid system also could be, in principle, the basis for long-term anti-Hebbian unlearning. However, more research must be done about this point since what we need is not blocking of long term memory formation, but weakening of already consolidated memory traces. We suggest that, when searching for such anti-Hebbian phenomena, it would be preferable to study *associative* LTP/LTD, that is, synaptic changes dependent on the temporal order between pre and post-synaptic depolarization (Bi & Poo 2001) instead of the usual tetanus induced LTP/LTD.

At the behavioral level, the ansiolitic and (indirect) rewarding properties of cannabinoids could function as a positive context to be associated with the fear cues present in the dream. The organism would learn to associate previous fear cues with non-aversive outcomes, like in desensitization therapy (Marks & Tobena 1990). For example, the instinctive human fear of falling can turn out obsessive (acrophobia). Falling dreams and flying dreams, associated with cannabinoid effects, could provide extinction of this fear, like the extinction of fear observed in people addicted to role-coasters or parachuting.

So, we can elaborate a strong and a weak unlearning hypothesis: in the strong hypothesis, true anti-Hebbian unlearning weakens the part of the synaptic matrix that stabilizes strong patterns and associations (for example, fear-learning responses). However, such exact anti-Hebbian mechanism needs yet to be found experimentally. In the weak hypothesis, we have usual extinction phenomenon, that is, association of fear cues to a non-aversive (or even rewarding) outcome elicited by endocannabinoids and opioids liberated during REM-sleep.

## V. Some predictions and experimental tests of the unlearning hypothesis

Our two related hypothesis for REM-sleep/dream function (neurological role = weakening of over stable attractors, behavioral role = emotional habituation performed during simulated exploratory behavior), beyond providing a more inclusive account of the experimental data than the MCH, has indeed strong predictive power. We discuss in the following some general predictions and some specific experimental suggestions.

**Prediction 1**: **Extinction of fear learning is made at the amygdalar complex during REM-sleep/dreams.**

Fear learning extinction should be more effective after REM-sleep periods. The neurochemical agents that mediate the extinction process should be produced during REM-sleep. In particular, we proposed that endocannabinoids mediate synaptic weakening and unlearning in the baso-lateral amygdala during REM-sleep. Their presence should be experimentally detectable, perhaps by following the methods of Murillo-Rodriguez (1998; 2001; 2001b).

**Prediction 2**: **There should be spatial correlation between the CB1 receptor density and plastic regions highly activated during REM-sleep.**

This correlation is already visible comparing a list of regions that densely express CB1 receptors with a list of regions that are strongly activated by REM-sleep. But a finer spatial analysis could be done, for example by using the amygdalar CB1 density measured by Katona et al. (2001).

**Prediction 3**: **Dream amnesia reflects the unlearning process and is mediated by endocannabinoids.**

Dream amnesia is a natural effect under the Unlearning Hypothesis but seems to be misplaced in the context of Memory Consolidation Hypothesis. Contrasting to MCH, anti-correlation between REM density and dream recall should be expected: less REM means less endocannabinoids and better memory formation. Preliminary evidence of such phenomena has been reported by Pace-Schott et al. (2001). One also could test if cannabinoids antagonists produce better remembering of dream content.

**Prediction 4**: **REM-sleep and dreaming are not critical to cognitive performance.**

In contrast to MCH, our primary role for REM-sleep in the mature brain (control of anxiety and obsessive states, emotional unlearning) are not cognitive. Extensive REM-blockade (say, by phenelzine (Landolt & Boer 2001)) with mild cognitive impairments is conceivable under the Unlearning Hypothesis, since the drug substitutes REM-sleep in its presumed anti-obsessive properties. So, the apparently harmless phenelzine REM-blockade is strong counter evidence to MCH (Vertes & Eastman 2000) but seems to be not counter evidence at all to the Unlearning Hypothesis.

The postulated behavioral role for dreaming brings long term advantages to the organism, enabling the emergence of dream narrative by natural selection, but can be skipped during laboratory times without detectable loss. Notice the difference between critical biological functions (like memory consolidation) and functions not critical for short-term functioning as exploratory behavior. Exploratory behavior (real or simulated) brings emotional robustness and consequent survival advantage to organisms, but temporary absence of exploratory activity is not very much damaging.

**Prediction 5**: **Excessive REM-sleep impairs memory consolidation.**

Although REM-sleep deprivation probably is not so harmful, REM-sleep excess could impair memory by unlearning what should not be unlearned. This is a clear prediction of the computational model. It is also known that excess of cannabinoids impairs memory. Memory consolidation in rats has been impaired after REM-sleep enhancement due to Anandamide (Murillo-Rodriguez et al. 1998; 2001) and Oleamide (Murillo-Rodriguez et al. 2001b).

We need to have some caution about this prediction, however, because the biological system probably has safeguards to such potential danger. For example, the CB1 receptors desensitize under high endocannabinoids concentration after two hours (Kouznetsova et al. 2002), so unlearning may be stopped by this receptor desensitization. Is this desensitization time that sets the need for several 90 minutes REM-sleep cycles interleaved with non-REM sleep for recovery of CB1 receptors?

**Prediction 6**: **REM-sleep is essential for developmental synaptic pruning.**

There exists recent evidence that developmental synaptic pruning occurs on synapses previously weakened by Hebbian processes (Colman et al. 1997). REM-sleep could be critical for such phenomena by performing this preliminary weakening. For example, it is known that the formation of ocular dominance (which depends on synaptic loss and axonal withdraw) is delayed by REM-sleep blockade (Marks et al. 1995). In our interpretation, this delay reflects the fact that synapses have not been properly weakened by REM-sleep.

The Unlearning Hypothesis predicts that one should observe temporal anti-correlation between the REM-sleep density curve and the synaptic density curve during the developmental process. If the cannabinoid hypothesis is also correct, then one should have correlation between the CB1 receptor density and massive pruning phenomena. The connection between REM-sleep and synaptic pruning could also be tested by examining the effect REM-sleep suppression during pruning epochs.

**Prediction 7**: **Highly plastic tissues like amygdala and hippocampus should turn out more prone to seizures in the absence of REM sleep.**

Seizures produce strong correlated Hebbian activity that should induce LTP, turning the seizure state a more stable attractor. If REM-sleep detects and weakens strong attractors, then it should have some role in the control of seizures. There is evidence of exacerbation of seizures after REM-blockade (Ehrenberg 2000). There is also some evidence that cannabinoids have anti-epileptic properties (Wallace et al. 2001). Notice that MCH, being a cognitive hypothesis, has nothing to say about REM-sleep and seizures.

**Prediction 8**: **Replaying games is done to unlearn them.**

Dreams as virtual exploratory behavior explains why some activities idiosyncratic to modern humans (for example, visual spatial games: chess, Tetris, video games) are easily induced in dreams and hypnagogic imagery after intensive training (Stickgold et al. 2000). Similar to dreams, these spatial games strongly activates hippocampal memory, visual cortex, motor cortex, basal ganglia and cerebellum. But, contrasting to MCH, we view these dream and hypnagogic imagery as regulatory mechanisms: repetitive and strong learning protocols induce excessive plasticity that must be reversed by unlearning (and endocannabinoids). We also notice that the chess or Tetris scenarios dreamed are not simple replays of recent experience but stereotyped experiences full of anxiety: the blocks do not fall at the right places; the chess positions configure aversive situations or difficult problems. So, we suggest that if we add to the learning protocol some aversive outcomes (like some punishment for errors in Tetris training) the probability of related hypnagogic imagery should be increased. This could be quantified by experimental measurements like those made by Stickgold et al. (2000).

**Prediction 9**: **Weak associations are strengthened and strong associations are weakened during REM-sleep.**

The computational model suggests that not only the stability of strong patterns (associations) is weakened, but also, at the same time, the stability of weak patterns (associations) is enhanced. Recently, such phenomenon has been detected by Stickgold and co-workers (1999): after REM-sleep, weak priming (association between low related words) is enhanced and strong priming (association between strongly related words) is weakened.

**Prediction 10**: **Dream acceleration should occur during REM-sleep.**

The unlearning procedure, by diminishing the stability of patterns, produces an *acceleration phenomenon*. This means that the frequency of transition between the meta-stable states grows with time. Notice that this is not a particular feature of the model, but it should be present in any itinerant dynamics where attractor stabilities are slowly weakened.

A similar phenomenon has been previously observed by Anderson and Kawamoto in a model with anti-Hebbian synaptic dynamics (Kawamoto & Anderson 1985). Anderson and Kawamoto linked the transient acceleration phenomenon presented by their model with the psychological acceleration observed in visual bistability phenomena. This transient acceleration is due to the accumulation of the anti-Hebbian terms until a stationary state is achieved. Here we make a similar claim, and suggest that the so called *dreaming acceleration* is due to the increasing facility for escaping weakened patterns.

Replay of rat hippocampal ensembles is done at a faster pace in non-REM sleep (McNaughton 2000). We predict that the same phenomenon should occur during REM-sleep and that slower replay never shall be observed. Suggestion: recently Nicolelis and co-workers have implanted arrays of hundred of electrodes in the motor cortex of mammals and obtained strong correlation between collective firing patterns and animal movements (Nicolelis 2001; Kralik et al. 2001). The spatio-temporal pattern coding advocated by Nicolelis is similar to the used in this work. We predict that if REM-sleep is studied in Nicolelis experimental setup, one should observe a variety of meaningful motor patterns (full simulations, not simple replays). We also predict that these patterns might be activated at a faster pace at the final of the REM-sleep cycle.

## VI. Summary and Conclusion

This work intended to be an interdisciplinary study, an intellectual collaboration between a biological physicist (OK) and a psychologist (RRK). Our primary intention is not to present new experimental results but to draw a framework that reorganizes the known facts and suggests new directions for research. Our main contributions are summarized as follows:

**a) An improved computational model for implementing Crick-Mitchison's ideas.**

To our knowledge, anti-Hebbian synaptic regulation has not been explored as a driving mechanism in itinerant memory models. Our James Machine, after recalling the immediate pattern induced by the inputs, starts to recall related patterns, wandering and exploring the associative space. In the absence of external inputs, this itinerant process is the dominant dynamics of the network ("the dreaming state"). The coupling of such chaotic associative dynamics with a long term anti-Hebbian mechanism furnishes a robust and efficient way to control over imprinted patterns. The model solves an important weak point of Crick-Mitchison original formulation (absence of dream narrative), is compatible with the idea that "dreaming is a royal road to emotionally charged processes" and produces novel qualitative predictions.

Itinerancy based in anti-Hebbian synaptic dynamics (the James Machine) also seems to be superior to Horn and Adachi-Aihara models because it eliminates the fast oscillations between patterns and anti-patterns. The model predicts not only that strong associations are weakened, but also that weak associations (not visited during the dream) are enhanced, like recently found by Stickgold et al. (1999). Another important prediction is the acceleration phenomena that occur during unlearning: it gives a mechanistic account of acceleration in dreams and faster replay of hippocampal cells (McNaughton 2000).

**b) An improved hypothesis for REM-sleep function.**

Our neurological hypothesis about REM-sleep conceives it as a part a homeostatic circuit that weakens strong attractors in plastic brain tissues. At the developmental level, unlearning might be critical to massive synaptic pruning mechanisms that shape the mature brain. At the behavioral level, unlearning strong attractors would reflect in homeostasis of anxiety and control of obsessive states and other stereotyped brain activity.

### c) A neurobiological hypothesis for the unlearning mechanism.

We also observed that the cannabinoid system seems to be a prime candidate to furnish the neurobiological basis both for the itinerant dynamics and the unlearning process. Short-term suppression of synaptic efficiency by endocannabinoids is similar to our fast anti-Hebbian factor. A long term unlearning mechanism is conceivable given the known memory modulation properties of endocannabinoids. The connection between the cannabinoid system and REM-sleep is known and this interaction could be enhanced by the sleep inductor cannabinoid Oleamide. We recognize, however, that the specific cannabinoid-based unlearning mechanism needs yet to be fully formulated.

### d) An improved evolutionary hypothesis for dream narrative emergence.

Our evolutionary hypothesis about the emergence of dream narrative is that, by recurrent exposition to simulated aversive scenarios, mammals (and birds) improve their emotional response to aversive situations without exposing themselves to the risks present in real scenarios. This hypothesis is similar to Revonsuo's Threat Simulation Theory and several arguments for its plausibility can be found in Revonsuo writings. Contrasting to Revonsuo, we stress the weakening (unlearning) of fear, anxiety and other emotional responses, instead of consolidation (learning) of survival skills.

Hobson and McCarley tried to correlate features of dream imagery with spontaneously activated regions during REM-sleep. Here we try to answer why these specific regions and not others have been selected by evolution to be activated in the REM-sleep process. The evolutionary/behavioral role connects with the neurological role because circuits linked to exploratory behavior (locomotor system, cerebellum, hippocampus, amygdala) are the most susceptible to excessive plasticity.

We feel that a strong point of our proposal is that it rescues dream narrative from its conceptual isolation. Instead of an idiosyncratic, weird phenomenon, dreams would be integrated into a larger set of animal behaviors like exploration and play. We suggest that, since in our theory play and dreams perform similar functions, one could search for phylogenetic correlation between time spent in dreaming and time spent in playing.

Indeed, it is possible to suggest that a primary drive for some recent cultural inventions (bedtime stories, literature, theatre and movies) is that they are dream-like, that is, mimic the habituation of emotional responses  (and perhaps evoke the same endocannabinoid recompenses) already performed by the biological process of dreaming.  Notice that, in some cultures, telling dreams occupy the same cultural niche of telling stories and myths. Even the infantile expectancy for happy ends after the characters confront aversive and emotionally charged situations may reflect that dream-like biological motivation, to despair of artistic film makers. "Cinema as a fabric of dreams" (Goddard) would not be a superficial analogy: paraphrasing Hobson and McCarley, movies could be defined as *a mental experience, occurring in the dark, which is characterized by hallucinoid imagery, predominantly visual and often vivid; by bizarre elements due to such spatiotemporal distortions as condensation,*

*discontinuity, and acceleration; and by a temporary acceptance (suspension of disbelief) of these phenomena as "real" at the time that they occur.* We wonder if brain scanning images of subjects watching (action or terror) movies would reveal dreaming-like patterns of brain activity as high activation of amygdala and low activation of pre-frontal lobes.

**e) An improved reevaluation of James dynamical ideas.**

Modern advocates of the dynamical approach in Psychology are tributary to James tentative of rescuing the study of thought processes from simplistic accounts based in atomic symbol manipulation. His fundamental idea that a better analogue to the thought flux is a complex stream is surprisingly contemporaneous (Freeman 2000; 2001). Moreover, James emphasis on the multitude of unstable equilibria and punctuated transitions, characteristic of dynamical systems with many degrees of freedom and several time scales, illustrate a more proper framework to model the thought flux than approaches naively inspired in low-dimensional deterministic chaos (van Gelder 1998).

Our James Machine is a physical extended system with many degrees of freedom where macroscopic patterns of activity appear only to lead to the next ones in an incessant flux. Pursuing James stream analogy, one could remember that the complexity of actual river basins is due to two competing factors. A strong flux produces erosion that creates facilitated channels to follow (a "learning" effect). If this kind of positive feedback were the unique important factor, the river would follow its same obsessive ways again and again. But erosion also leads to the elevation and weakening of deep channels (an "unlearning" process), creating all the meandering complexity of a river basin. With Crick and Mitchison, we suggest that REM-sleep and dreams are related to a similar homeostatic mechanism in brain plasticity. Although the proper concepts could not be used in a rigorous way by James, his scientific intuition and literary skill enabled him to state in today easily recognizable terms that it is the combination of chaotic itinerancy and long term plasticity that prevents our Herakleitian thoughts from recurring twice.

**VII. Acknowledgements:** The authors acknowledge financial support from FAPESP and CNPq. O. Kinouchi thanks useful discussions with Sidarta Ribeiro, Antônio C. Roque da Silva, Nestor Caticha, Frederico Graeff, Silvia M. Kuva and Rodrigo F. de Oliveira.